\documentclass[journal]{IEEEtran}
\ifCLASSINFOpdf
\else
\fi
\usepackage{algorithm,algorithmic}

\usepackage{graphicx}

\begin{document}
%
\title{Investigating Underlying Drivers of Variability in Residential Energy Usage Patterns 
with Daily Load Shape Clustering of Smart Meter Data}

\author{Ling Jin, 
        C. Anna Spurlock, 
        Sam Borgeson, 
        Alina Lazar,
        Daniel Fredman, 
        Annika Todd,
        Alexander Sim,
        Kesheng Wu

\thanks{L. Jin, C. A. Spurlock and A. Todd are with the Energy Analysis and Environmental Impacts Division, Lawrence Berkeley National Laboratory, Berkeley, CA 94720 USA (e-mail: ljin@lbl.gov, 
caspurlock@lbl.gov, atodd@lbl.gov).}
\thanks{S. Borgeson is with Convergence Data Analytics, LLC, Oakland, CA}
\thanks{A. Lazar is with Youngstown State University, Youngstown, OH}
\thanks{D. Fredman is with University of Vermont, Burlington, VT}
\thanks{A. Sim and KS. Wu are with the Computational Research Division, Lawrence Berkeley National Laboratory, Berkeley, CA 94720 USA}}

\maketitle

\begin{abstract}
Residential customers have traditionally not been treated as individual entities due to the high volatility in residential consumption patterns as well as a historic focus on aggregated loads from the utility and system feeder perspective. Large-scale deployment of smart meters has motivated increasing studies to explore disaggregated daily load patterns, which can reveal important heterogeneity across different time scales, weather conditions, as well as within and across individual households. Such heterogeneity provides insights into household energy behavior and reveals sources and drivers of variability that is critical for utilities to understand in order to design efficient and effective demand side management strategies. This paper aims to shed light on the mechanisms by which electricity consumption patterns exhibit variability and the different constraints that may affect demand-response (DR) flexibility. We systematically evaluate the relationship between daily time-of-use patterns and their variability to external and internal influencing factors, including time scales of interest, meteorological conditions, and household characteristics by application of an improved version of the adaptive K-means clustering method to profile "household-days" of a summer peaking utility. We find that for this summer-peaking utility, outdoor temperature is the most important external driver of the load shape variability relative to seasonality and day-of-week. The top three consumption patterns represent approximately 50\% of usage on the highest temperature days. Having an electric dryer and children-in-home are the leading predictors of a more variable consumption schedule, while conversely homes with elderly residents exhibit the most stable day-to-day routines. Among the customer vulnerability characteristics considered here (chronic-illness, elderly, and low-income), we find low-income households tend to have more variable consumption patterns. The variability in summer load shapes across customers can be explained by the responsiveness of the households to outside temperature. Our results suggest that depending on the influencing factors, not all the consumption variability can be readily translated to consumption flexibility. Such information needs to be further explored in segmenting customers for better program targeting and tailoring to meet the needs of the rapidly evolving electricity grid. 

\end{abstract}


%
\IEEEpeerreviewmaketitle

\section{Introduction}
%
%
%
%

 
With the rise of residential Advanced Metering Infrastructure (AMI) in the past decade, increasing prevalence of high-resolution meter data has motivated more research to apply load profiling to residential customers (e.g. \cite{Rasanen2010-xl, Flath2012-xl, Cao2013-jz, McLoughlin2015-eo,Khan2019-cz, Kwac2014-rt,Haben2016-ma, Yi_Wang2015-ff, Wang2018-da}) that have traditionally not been treated as individual entities due to the high volatility in residential consumption patterns as well as a historic focus on aggregated loads from the utility and system feeder perspective \cite{Chicco2012-ol}. Data and evidence based insights into behavioral usage patterns hold the potential for efficient and effective load forecasting and planning, demand response management, time-of-use tariff design, and electricity settlement \cite{Moslehi2010-bs, Farhangi2010-nx, Hong2011-au, Zhou2013-ya, Wang2018-da}.

Despite this progress, most work have focused on load shape differences between customers and therefore applied clustering to individual households each associated with pre-averaged load shapes or attributes (e.g. averaged profiles by season, by day of week, or by month). Such an approach ignores the day-to-day variability within households, thereby overlooking fine-grained information on household attributes and sources of variation that could be valuable for subsequent classification models and segmentation. This omission can have meaningful repercussions in terms of our understanding of patterns of electricity load. A recent study by Yilmaz et al. \cite{Yilmaz2019-ra} has demonstrated significant differences in clustering results between daily load profiles averaged over individual households and raw daily load profiles (with no averaging); Kwac in \cite{Kwac2014-rt} found that although two homes might have the same average profiles, the diversity of load patterns from one day to the next could vary significantly; McLoughlin et al. in \cite{McLoughlin2015-eo} applied cluster analysis to day-to-day usage patterns of individual households and found the sequence of resulting daily patterns useful for further segmenting customers; and Haben et al. in \cite{Haben2016-ma} demonstrated day-to-day variability in energy consumption within the household together with their average behavior was sufficient to meaningfully distinguish households.

The diversity or variability of day-to-day time-of-use patterns within households is especially important in the context of demand response and energy efficiency programs, as such information may directly relate to each household’s suitability to various demand side management strategies. For example, it is speculated that households with variable consumption schedules may be more flexible and therefore likely to respond to time of use pricing incentives \cite{Yilmaz2019-ra}, whereas those with regular demand during the daytime are ideal to target for integrating solar energy \cite{dyson2014using}. However, these speculations have not been sufficiently empirically verified and specifically the underlying drivers of variability in day-to-day usage patterns are yet to be examined. Consequently there is no consensus in the literature on a comprehensive relationship between variability and flexibility (\cite{dent2014variability, Kwac2014-rt,Yilmaz2019-ra} and detailed in Section II).

In this study, we apply an efficient whole 24 hour time series clustering algorithm to each daily load profile (household-day) of a large sample of residential customers as opposed to aggregated load profiles for individual households. This approach allows us to assign each household to multiple representative load patterns that may vary from day to day, so that patterns in electricity consumption and their variability within and across households can be derived. We systematically examine this load shape variability based on distributional differences in the resulting dictionary of load shapes across underlying external (such as seasonality, day of week, outside temperature) and internal (such as socio-demographic and household properties) drivers. In contrast to previous studies which typically use load data sets ranging from a few hundred to a few thousand households (see review \cite{Wang2018-da}), our load data consists of more than 30 million daily load profiles from approximately 100,000 households.

The context of our analysis is a summer-peaking utility that has a time-of-use (TOU) electricity pricing option. This type of time-based pricing is an example of a mechanism, like demand response (DR) programs, designed to shift electricity consumption from the highest demand times of day through a monetary incentive. This shift relies on customers actively changing their behavior in response to these incentives. TOU pricing programs are gaining traction as attractive alternatives to more traditional flat or inclining-block electricity rates at the residential level. California, for example, has authorized their investor-owned utilities (IOUs) to institute default TOU pricing across their residential customer base \cite{Public_Utilities_Commission2009-qu}. Understanding underlying patterns of behavior at an individual household level and how these patterns relate to the timing relevant for a TOU program is therefore highly valuable.

Within this TOU pricing context, we focus our analysis specifically on customer-controlled electricity loads (e.g., lighting, air conditioning, computer equipment, entertainment, dishwashers, laundry equipment), rather than installed equipment demand (e.g., electric tank-type water heater or refrigerator load). Households are more able to quickly adjust these customer-controlled loads with active behaviors in a short-run response to a DR program like TOU pricing, relative to the more long-run response of replacing installed equipment with more energy efficient versions. To isolate this customer-controlled usage we introduce an additional innovation unique among the research in this area: we focus on clustering ``discretionary" electricity usage profiles (further defined in Section III B) rather than profiles of total hourly electricity use. 

The goal of this research is to demonstrate how daily consumption patterns and their diversity in discretionary electricity consumption within and across households can be explained by factors relevant to DR programs, such as day-of-week, season, meteorological conditions, and household characteristics. By doing so, we aim to shed light on the mechanisms by which electricity consumption patterns exhibit variability and the different constraints that may affect DR flexibility. The paper is organised as follows: Section II provides a review of clustering methods used in the literature and current understanding of load shape variability, thereby explaining the methods and approaches and defining the challenges. Section III presents the data and methods used in this study. The resulting dictionary load shapes and their distributional differences in relation to both external and internal factors are examined and discussed in Section IV. Section V concludes.

\section{Related work and our contribution}
\subsection{Direct load shape clustering}
Cluster analysis is a commonly used unsupervised learning technique used for load profiling that can help discover and understand patterns in electricity consumption. This study applies whole 24-hour time series clustering to daily load profiles, which falls under the direct-clustering based approach according to \cite{Wang2018-da}. As reviewed in Chicco \cite{Chicco2012-ol}, a number of direct clustering techniques, such as k-means, follow the leader, and self-organizing maps, were applied to whole-building load data to construct load profiles for non-residential (i.e., industrial and commercial) customers. Residential customers are characterized by highly volatile behavior, which challenges the application of clustering methods to individual load curves \cite{Chicco2012-ol}. Using a large sample of residential daily load profiles ($>$100,000) and six performance metrics Jin et al. \cite{Jin2017-dv}, following \cite{Dent2012-yu}, conducted a comparative study to evaluate eleven direct clustering methods under four families of algorithms: centroid based, hierarchical, density based, and model based methods. They found whole time series clustering of residential load profiles exhibits a trade-off between cluster compactness and distinctness and the number of clusters required to achieve adequate performance was 50 to 100, much larger than that of non-residential customers.

The key to data synthesis in this context using direct clustering is to identify a diverse set of typical daily shapes that can be adequately described by the cluster centroids representing different patterns in day-to-day and customer-to-customer consumption schedules. \cite{Jin2017-dv} found that algorithms with heuristics minimizing the within cluster scatter, such as k-means and adaptive k-means, perform better with respect to such load profiling goals. We improve upon adaptive k-means \cite{Kwac2014-rt} and explicitly control for clustering quality with reasonable statistical affinity of cluster centroids.  

As the interest in time-of-use patterns is generally and primarily the temporal aspect of the daily profile rather than absolute usage, the load profiles are usually preprocessed with normalization. Most existing studies normalize the daily usage data by a reference power value following standardizing methods reviewed in Milligan and Cooper \cite{Milligan1988-ql}. For example, Chicco et al. \cite{Chicco2006-jz} and Chicco \cite{Chicco2012-ol} divided hourly usage by the daily maximum; Piao et al.\cite{Piao2014-gu}, Han et al. \cite{Han2011-ev}, and Cao et al. \cite{Cao2013-jz} employed min-max normalization, which subtracts the minimum from the data and divides by the maximum; and Kwac et al. \cite{Kwac2014-rt} normalized hourly demand by the daily total. We also apply a normalize-to-one procedure, however, prior to this step we isolate discretionary electricity consumption from total hourly consumption through a “de-minning” process described in Section III.

\subsection{Load shape variability and demand-side flexibility}
Demand-side flexibility is defined as the ability for consumers to change how, when, and where energy is used \cite{torriti2015peak}. As the energy system becomes increasingly supplied by variable renewable generation, predictable and/or controllable flexibility becomes increasingly important for balancing supply and demand \cite{satre2019daily}. The variability of the daily load "shapes" or the day-to-day changes in consumption schedules of households provides a reasonable indication of whether and how the households may change their energy usage in response to a utility program \cite{Yilmaz2019-ra}.  Note that this paper is focused on variability or diversity in the consumption schedules, i.e. the "shape" of daily loads, rather than variability in absolute electricity consumption at given times of the day (e.g. \cite{Khan2019-cz,Haben2016-ma}) or the intra-day variations in consumption levels (e.g. \cite{Roberts2019-ck}), which are also important features to explore. 

Past studies have employed load shape clustering to explicitly quantify the "shape" variability \cite{Xu2017-lt, Zhou2016-ec, Kwac2014-rt} as a measure of potential demand response flexibility. However, there has not been consensus in the literature on a comprehensive relationship between variability and demand response flexibility. Using a data set collected from a residential demand response program, \cite{Zhou2016-ec} found that customers with more variable consumption patterns are more likely to reduce their consumption compared to those with a more regular consumption behavior. In contrast, \cite{Kwac2014-rt} proposed that a more stable household that shows the same load shape every day should be targeted by DR programs rather than one that is highly variable. As reviewed by \cite{satre2019daily}, there are significant literature gaps in both the mechanisms by which electricity consumption patterns exhibit variability and the different constraints or motivating factors that can reshape them. 

Emerging studies have taken a deeper dive into the underlying drivers of load shape variability including season, day of week, temperature, and household characteristics. \cite{Fang2018-xc} used a Hidden Markov Model to learn the consumption dynamic behavior under the corresponding environments and concluded that customers can be grouped in three categories: normal, sensitive, and insensitive households in relation to outdoor temperature changes. \cite{Chen2017-gy} applied agglomerative hierarchical clustering based on proportions of different load curve categories in different seasons and found the behavioral patterns of customer groups are highly consistent across several seasons. \cite{Melzi2017-re} used a constrained Gaussian mixture model whose parameters vary according to the day type (weekday, Saturday or Sunday), and crossed the clustering results with contextual variables available for the households to show the close links between electricity consumption and household socio-economic characteristics. 

\cite{Rhodes2014-pu, McLoughlin2015-eo} correlated load profile classes with socio-economic determinants yet did not explore the variability linkage. \cite{economics2012demand} indicated that understanding DR potential of vulnerable and low-income customers was especially lacking. \cite{cappers2018vulnerable} found no evidence that vulnerable populations (low income, elderly, or chronically ill) were unduly harmed or burdened by a time-of-use pricing program, but did nothing to characterize the usage of such customers, which could provide an better understanding of the mechanisms through which such customers respond, or not, to that or similar programs. Using the dictionary of household-day load shapes derived from our study, we are able to systematically examine the correlation between load shape variability across various temporal scales and customer characteristics (including the above mentioned vulnerability characteristics) to better understand the source of variation and their implications for DR flexibility.

Lastly, entropy has been a common metric to quantify load shape variability and its application has been limited to characterizing individual customers \cite{Xu2017-lt, Zhou2016-ec, Kwac2014-rt}. Essentially, entropy quantifies the distribution or diversity of a given set of load shapes. In this paper, in addition to computing load shape entropy of individual customers, we also use this metric more flexibly to characterize load shape variability under different time scales, days, and outside temperature ranges to understand the role of these external drivers. 

\begin{figure*}[!ht]
\begin{center}
  \includegraphics[width=.78\textwidth]{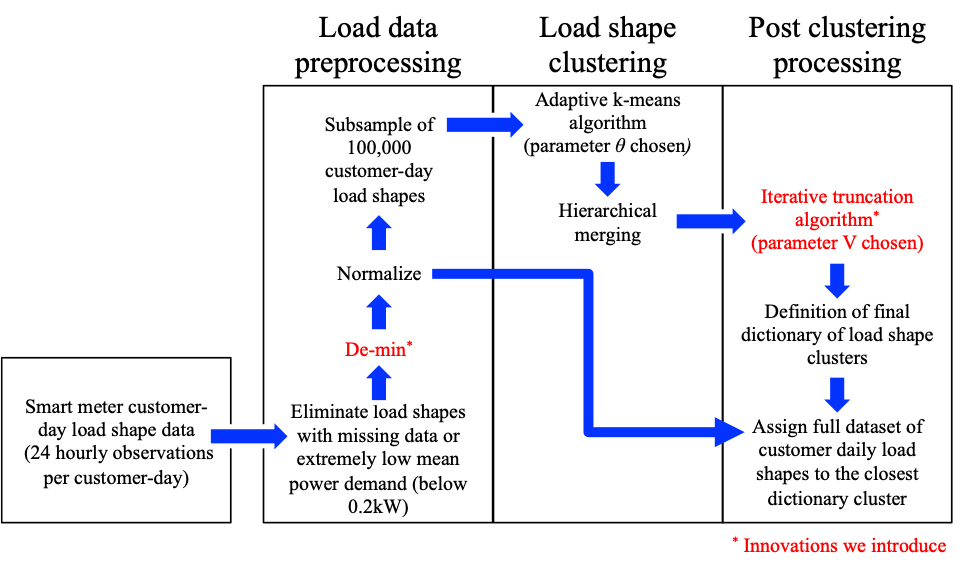}
  \caption{Modified adaptive k-means procedure to derive representative dictionary load shapes and their membership based on discretionary usage patterns}
  \label{fig:fig1}
  \end{center}
\end{figure*}

\section{Materials and methods}
\subsection{Dataset}
We cluster household-day load profiles based on hourly consumption data collected from a summer-peaking utility in California. The data consist of over 30 million daily load profiles (“household-days”) from approximately 100,000 households, measured between June 1st, 2011 and May 31st, 2012, prior to the implementation of time-of-use (TOU) pricing. Therefore, the data set represents consumption behavior absent any time-based rate or other related program. 

In addition, household information was collected in a survey of 6413 participants in the utility service area. The household characteristic variables used in this study are processed into binary indicators on: (1) socio-demographic and lifestyle information (low-income, chronically-ill, elderly, children-in-home, college-degree, work-full-time, work-from-home); (2) dwelling information (single family home); (3) appliance ownership information (electric dryer, central air conditioner, room air conditioner, programmable thermostat). Note that we explicitly included three vulnerability indicators (low-income, chronically-ill, and elderly) in order to inform the literature gap identified in Section II B.  

The clustering method employed here improves upon the method developed in \cite{Kwac2014-rt} and is illustrated in Figure \ref{fig:fig1} and described in detail below.

\subsection{Deriving discretionary load shapes}
We first conduct data cleaning and establish the format of the object to be clustered. We apply the same preprocessing criteria as Kwac et al. \cite{Kwac2014-rt}: dropping daily usage data with missing hour observations, or with low average demand (below 0.2kW). These data cleaning steps result in 32,611,421 daily load shapes (94\% of the raw data) remaining. From this set of cleaned load shapes a random subsample of 100,000 as suggested by Kwac et al. \cite{Kwac2014-rt} was drawn for purposes of clustering.

In the original adaptive k-means study, Kwac et al. \cite{Kwac2014-rt} normalized hourly usage by its daily total, so the area under each single household’s 24-hour load shape is one. We also apply a normalize-to-one procedure, however, prior to this step we isolate discretionary electricity consumption from total hourly consumption through a “de-minning” process described below. We then normalize by dividing each hourly discretionary usage value by that day’s total discretionary usage.

A household’s “discretionary” usage captures the electricity consumption resulting from active residential behavior (e.g., lighting, air conditioning, computer equipment, entertainment, dishwashers, laundry equipment). We innovate beyond Kwac et al. \cite{Kwac2014-rt} to isolate only discretionary consumption by “de-minning” the load profiles prior to normalization. Specifically, the daily minimum electricity usage is subtracted from each hour of that day within each household-day profile. The object to be clustered is therefore defined to be this “de-minned” and normalized profile of discretionary daily usage. 

This procedure has two advantages: first, from a conceptual perspective daily minimum electricity usage serves as a proxy for “baseload” so this procedure allows us to isolate a household’s variable, or discretionary, usage from their baseload. After normalization, a load shape essentially represents a sequence of each hour’s proportional contribution to that day’s total discretionary usage, and dictionary load shapes can be interpreted in terms of the overall patterns in timing of higher and lower discretionary use. 

\begin{figure}[ht]
\centering
  \includegraphics[width=.98\columnwidth]{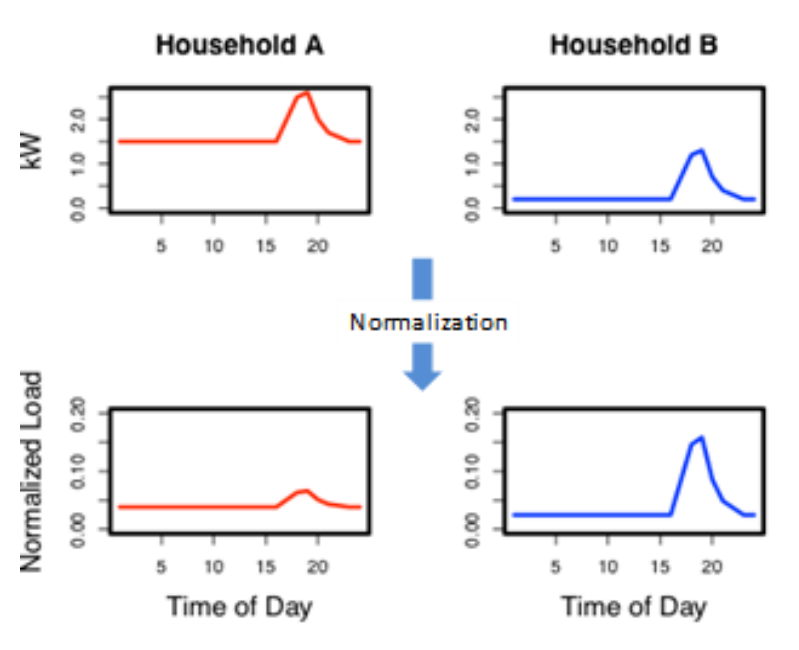}
  \caption{Illustration of the flattening effect of normalization without de-minning}
  \label{fig:fig2}
\end{figure}

The second advantage to this “de-minning” process is that it alleviates the distortion of consumption profiles that occurs during normalization when using the total daily electricity load for each household. In particular, a load profile with high baseload tends to be flattened when total hourly usage is divided by daily total consumption in the normalization step. To demonstrate this, the top row of Figure \ref{fig:fig2} illustrates two load shapes with the same discretionary consumption schedules but different baseloads, and the bottom row shows those same shapes after normalization without “de-minning”. This figure demonstrates that when the daily load has not been “de-minned” the normalization step causes the signal associated with the relevant variation in electricity usage stemming from the same active consumption behaviors to be significantly muted when there is high baseload and not muted when there is very low baseload. Subsequently moving to the clustering step when this is the case tends to result in one of the resulting representative dictionary clusters having a large membership consisting of undifferentiated flattened load shapes due to the nature of the distance metrics used to score shape fits into best-fit clusters. This means that any information regarding patterns of discretionary electricity consumption behavior is obscured. 

By “de-minning,” we significantly reduce this problem\footnote{Specifically, clustering the load profiles after normalization without “de-minning” resulted in more than 65\% of the daily load profiles in the data being assigned to a single flat-shaped cluster. When the daily load profiles were “de-minned” prior to normalization and clustered into the same number of clusters, the highest concentration of load profiles assigned to a single cluster was approximately 10\%.}. 

\subsection{Load shape clustering with adaptive k-means}
The application of adpative k-means to our dataset is more thoroughly documented in an earlier report \cite{jin2016load} and can be found in the supporting materials and briefly described here. After preprocessing, the subsample of 100,000 “de-minned” and normalized load shapes is moved into the load shape clustering step. They are first passed through an adaptive k-means\footnote{akmeans: Adaptive Kmeans algorithm based on threshold. R package version 1.1. https://CRAN.R-project.org/package=akmeans} algorithm (\cite{Kwac2014-rt}), which splits the data set of load shapes into $K_1$ clusters, such that the relative squared error (RSE) of any load shape assigned to a cluster is not greater than an error threshold $\theta$. The RSE is defined in the equation below, where $s$ is the load shape of interest, $t$ is the hour of day index, and $C_i$ is the cluster center to which $s$ is assigned. The error threshold $\theta$ is varied from 0.05 to 0.5 to determine a suitable value that results in the most reasonable $K_1$.

\begin{equation}
    RSE_{s,i} = \frac{\sum_{t=1}^{24}(s(t)-C_i(t))^2}{\sum_{t=1}^{24}(C_i(t))^2}
\end{equation}

As the resulting clusters from adaptive k-means are typically highly correlated, in the second step we follow Kwac et al. \cite{Kwac2014-rt}, and undertake a subsequent hierarchical merging of the clusters by sequentially combining the most similar clusters until their total count reaches a target number $K_2$. Under this transformation the requirement that all RSEs fall under $\theta$ is relaxed. In particular, the target size $K_2$ is selected such that it is the smallest number of clusters for which less than 5\% of the load shapes violate the $\theta$ threshold condition. Guided by the acceptable error threshold, we allow our original number of clusters to grow into the thousands before applying quantitative criteria in the post clustering processing step described in the next section.

\subsection{Iterative dictionary truncation}
In the post clustering processing phase, we implemented an iterative truncation algorithm (illustrated in Algorithm 1) to remove cluster centers with low member counts with a user-defined clustering quality metric: the overall violation rate ($V$: the fraction of load shapes with RSE $>$ $\theta$). Dictionary truncation allows us to focus on the electricity consumption patterns that represent the majority of the household-day observations, rather than outliers. In contrast to the original adaptive k-means algorithm, the parameter $V$ we introduced here ensures the maximum reduction of final dictionary size while controlling for clustering quality. The advantage of this post-processing procedure is to avoid tuning of the number of clusters $K$ (a usual hyper-parameter in k-means-type of clustering) because an optimal $K$ is automatically determined after the iterative truncation. Without this iterative process (i.e. the original adpative k-means implemented by \cite{Kwac2014-rt}), the clustering quality as indicated by the violation rate generally increases by 30

 \begin{algorithm}
 \caption{Iterative dictionary truncation.}
 \begin{algorithmic}[1]
 \renewcommand{\algorithmicrequire}{V:}
 \renewcommand{\algorithmicensure}{\textbf{$\theta$:}}
 \REQUIRE violation rate
 \ENSURE  error threshold
 \\ \textit{LOOP Process}
  \WHILE {violation $<$ V}
  \STATE Identify the ids of smallest clusters whose shape members comprise the fraction V of the total number of shapes
  \STATE Remove those clusters
  \STATE Reassign the shapes that were members of the removed clusters into the remaining closest clusters
  \STATE Compute violation rate as fraction of load shapes with RSE $>$ $\theta$
  \ENDWHILE
 \end{algorithmic} 
 \end{algorithm}

Following the truncation procedure the resulting set of remaining cluster centers are defined as the “dictionary” of discretionary usage patterns (referred to as “dictionary load shapes”). Finally, each household-day in the full data set is assigned to the single closest dictionary load shape based on Euclidean distance.

\subsection{Entropy to quantify load shape variability}
We use load shape entropy (defined below) to quantify the distribution or diversity of a given set of the load shapes. Greater entropy indicates the load shapes are distributed more uniformly and therefore more diverse and variable, while smaller entropy indicates the distribution is concentrated on fewer dictionary load shapes and is therefore less variable. 

\begin{equation}
    \mathbf{S}_i = - \sum_{c \in C_i}{p_{i,c} \bullet log(p_{i,c})}
\end{equation}

Where $\mathbf{S}_i$ is the Entropy of set $i$; $C_i$ is the set of dictionary load shapes observed in set $i$; $c$ is any dictionary load shape that occurs within set $i$, and $p_{i,c}$ is the frequency of dictionary load shape $c$ that occurs within set $i$.

The set $i$ is usually defined by customer, that is, $C_i$ is the set of dictionary load shapes observed for customer $i$ over a certain time period. In addition, to understand the overall relationship between variability and external factors, we also compute population-level entropy of a given temporal period. In such a case, the set $i$ is defined by season, day-of-week, days with a certain temperature levels, or by each day. For example, $C_i$ could be the set of dictionary load shapes observed for the summer season across all customers.

\section{Results and Discussions}
\subsection{Clustering results and descriptive analysis}
\begin{figure}[ht]
\centering
  \includegraphics[width=.78\columnwidth]{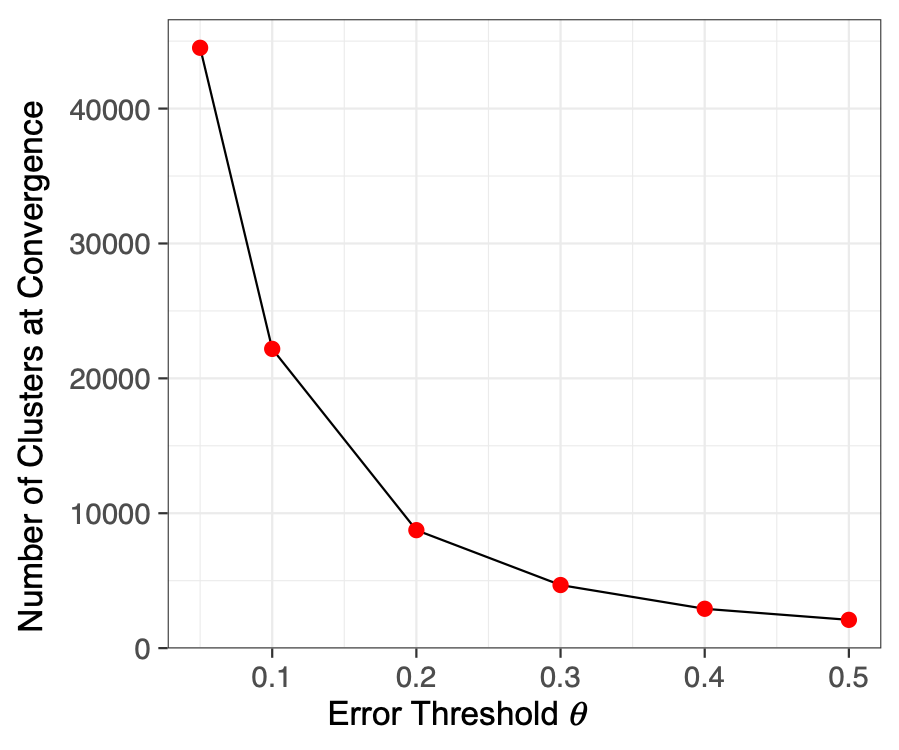}
  \caption{Error threshold ($\theta$) and number of clusters}
  \label{fig:fig3}
\end{figure}

The clustering results were documented in a previous report \cite{jin2016load} and briefly described here. The resulting number of clusters from the adaptive k-means procedure depends on the chosen error threshold ($\theta$). The relationship derived by running adaptive k-means for $\theta$ varying from 0.05 to 0.5 is plotted in Figure \ref{fig:fig3}. We selected $\theta$ = 0.3 based on our criteria that the number of clusters compared to the original number of household-days should not be excessively large (approximately 5000 initial clusters in our case), and the marginal gain in error improvement to the explanatory power by increasing $\theta$ should be small.

By limiting the total share of load shapes that violate the $\theta$ threshold to 5\% as suggested by Kwac et al. \cite{Kwac2014-rt}, the hierarchical clustering step consolidated the number of clusters from approximately 5000 ($K_1$) to 2000 ($K_2$) in our case. The top 600 of these 2000 clusters (sorted by member count) account for approximately 90\% of the data. It is because of this long tail of low-membership clusters that we applied the truncation procedure described in Algorithm 1. We use the iterative truncation algorithm with violation rates ($V$) of 10\% and 30\%, which results in sets of cluster centers numbering 608 and 99, respectively. We evaluate these two sets of representative clusters by using the Davies-Bouldin index (DBI), which is a metric of cluster separation (\cite{Davies1979-wh, Dent2012-yu}). The DBI for the 608 and 99 cluster center sets are 2.23 and 2.22, respectively, indicating both sets of clusters have similar performance with respect to cluster separation. Because of the increased ease in further analyzing a smaller set of clusters, we chose the 99 cluster centers as the final dictionary of representative load shape clusters to which the full data set of discretionary clusters are assigned based on smallest Euclidean distance.

\begin{figure}[ht]
\centering
  \includegraphics[width=.98\columnwidth]{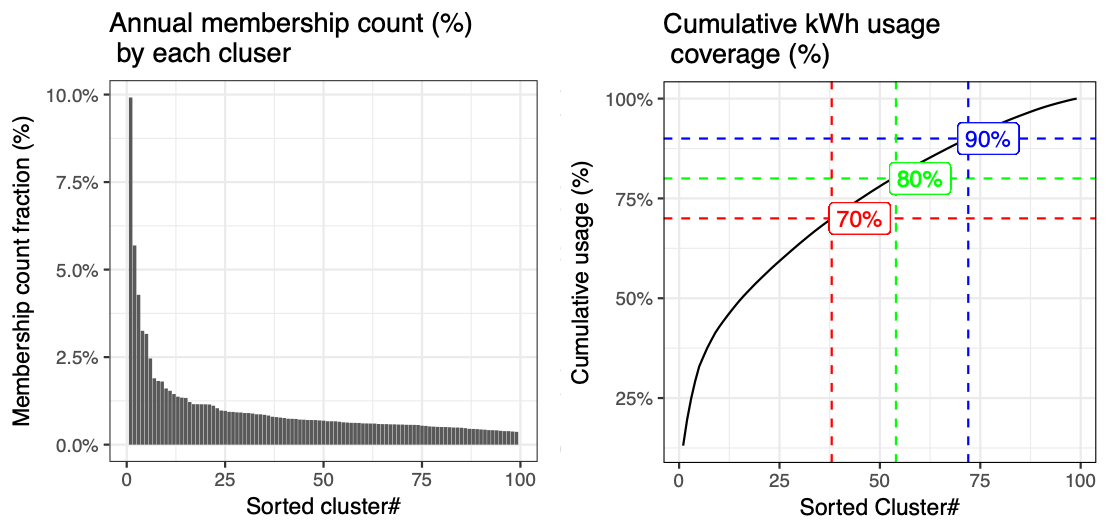}
  \caption{Distribution of dictionary load shapes (left) and their cumulative electricity coverage (right).}
  \label{fig:fig4}
\end{figure}

The distribution of kilowatt-hour (kWh) usage represented across all the 99 dictionary load shapes is not uniform; as is demonstrated in Figure \ref{fig:fig4}, the top cluster with respect to kWh of usage captured by that cluster accounts for approximately 13\% of the total kWh usage; the top 38, 53, and 73 dictionary load shapes respectively cover 70\%, 80\%, and 90\% of the electricity use underlying the approximately 30 million load shapes across households over the entire year period. 

\begin{figure}[ht]
\centering
  \includegraphics[width=.98\columnwidth]{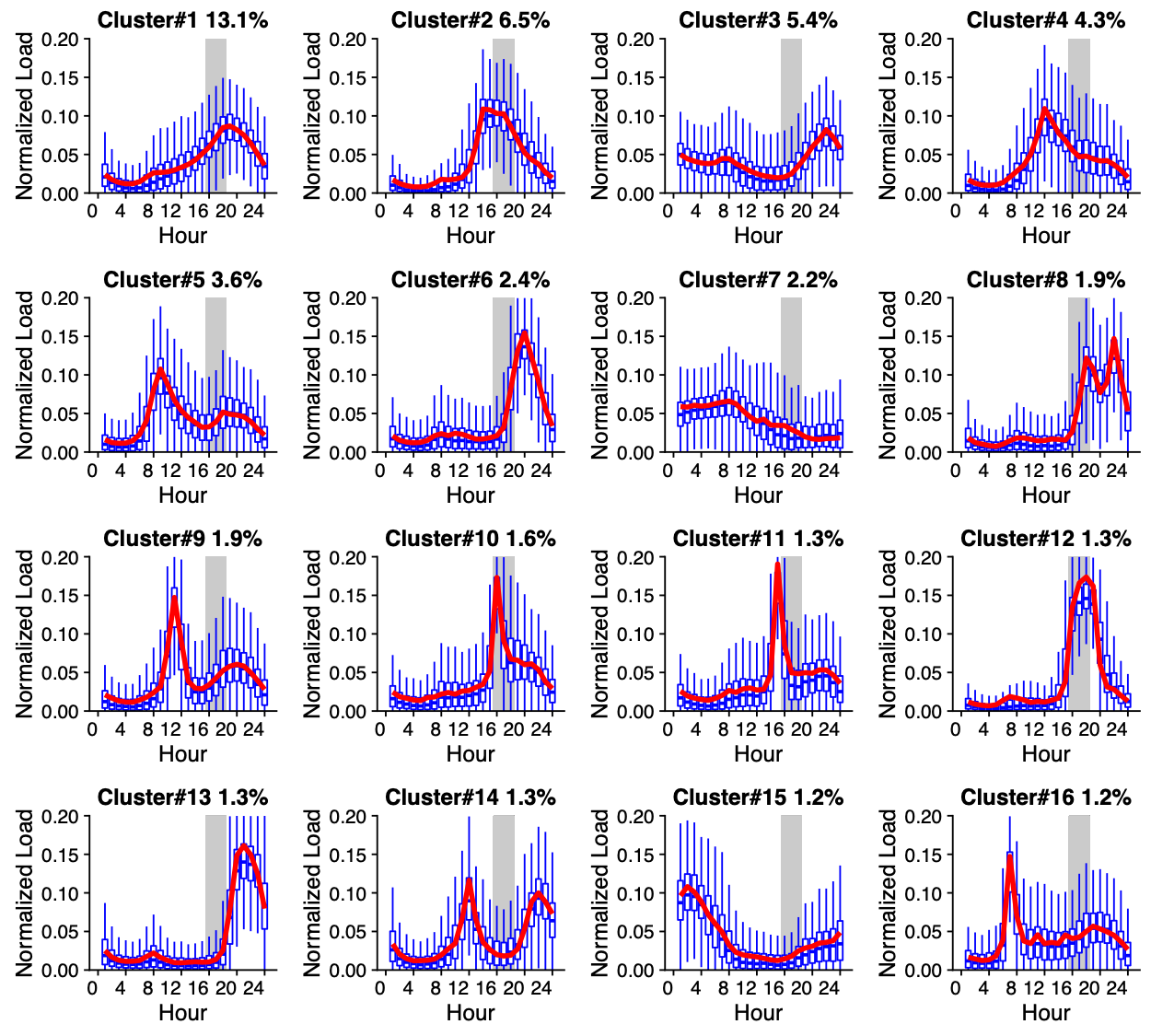}
  \caption{Sixteen highest electricity using dictionary load shape cluster centers. Notes: The title of each panel is: [Cluster\#] [Percentage of total daily electricity load covered by this cluster], where “Cluster\#” is the label assigned to that dictionary cluster, ordered based on the total kWh usage of its members. The boxes and whiskers summarize the within-hour distribution of load members belonging to the cluster (median, inter-quartile range, and 5th to 95th percentile), with the mean normalized discretionary load marked in red. The gray shaded area indicates the peak period for the TOU rate.}
  \label{fig:fig5}
\end{figure}

The top 16 dictionary load shapes of this final dictionary ranked by their total daily electricity load coverage are shown in Figure \ref{fig:fig5} and account for more than 40\% of all the household daily load shapes in the data set, and more than 50\% of total daily electricity load. The utility defines their peak period for the TOU rate from 4 to 7 PM (i.e., hours 16 through 18 as indicated by the gray shaded areas in Figure \ref{fig:fig5}) on non-holiday weekdays. While in aggregate, most of the high electricity usage happens during this period, Figure \ref{fig:fig5} demonstrates that the clusters exhibit considerable variability in peak timing of discretionary usage as well as number of peaks.  For example, the clusters with ranks 3 through 9 and 13 through 16 in Figure \ref{fig:fig5} reflect patterns of significant discretionary usage peaks that are outside of the TOU peak period. Conventional expectation of the most common load shape may be a morning peak and an evening peak, before and after work, but only clusters 5 and 16, representing only 3.6\% and 1.2\% of total daily consumption, respectively, really follow this classic pattern. Other double peaking load shapes follow different peak timing from the conventional expectation: for example, the second peak in cluster 3 occurs in late evening, while clusters 9 and 14 have distinct double peaks at noon and late evening.

\begin{figure*}[ht]
\begin{center}
  \includegraphics[width=.78\textwidth]{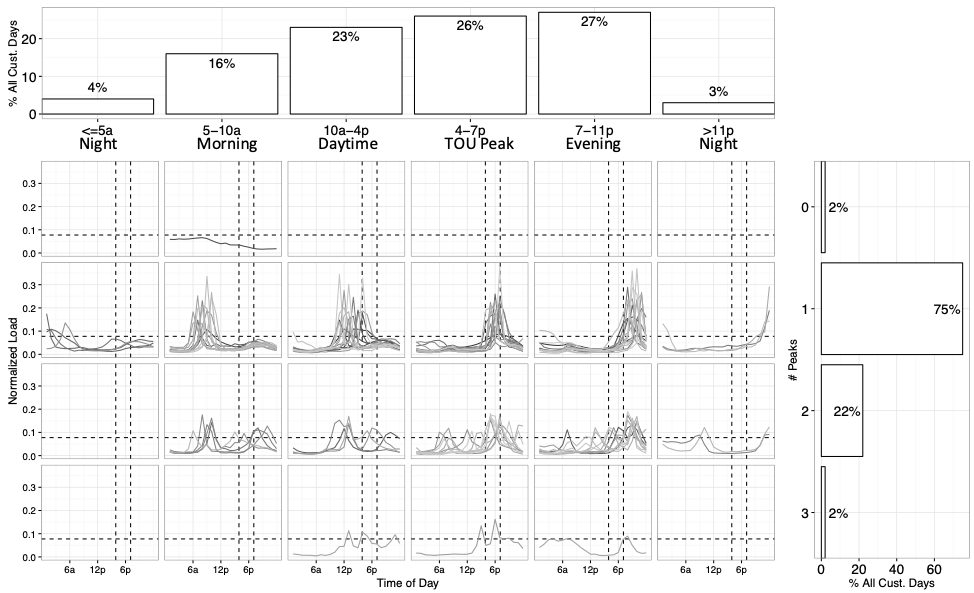}
  \caption{Dictionary load shapes categorized by peak timing (horizontal axis) and number of daily peaks (vertical axis)}
  \label{fig:fig6}
  \end{center}
\end{figure*}

Figure \ref{fig:fig6} aligns the 99 dictionary load shapes on a two dimensional space governed by peak timing (x-axis) and number of peaks (y-axis). Interpreting patterns within this figure broadly, we see that of all the household-day load shapes in the full year data set, approximately 75\% are assigned to dictionary load shapes that are single peaking, and approximately 22\% are double peaking.  Daytime (10 AM to 4 PM), TOU peak (4 to 7 PM), and evening (7 to 11 PM) are the most frequent times when major peaks occur, accounting for 23\%, 26\% and 27\% of the full year data set, respectively. 

Of particular interest in this set of results is the degree to which the system peak (defined as the TOU peak from 4-7pm) does not necessarily represent the most frequent peak in discretionary usage as represented by this dictionary of load shapes. Only 26\% of all household-day load shapes are assigned to dictionary load shapes that exhibit significant discretionary usage peaking during that time period. For this reason it is important to understand the underlying behavior generating these patterns to be able to best target and affect the behaviors that may have the largest impact on the system peak.

\subsection{Temporal and meteorological heterogeneity}
\begin{figure}[ht]
\begin{center}
  \includegraphics[width=.98\columnwidth]{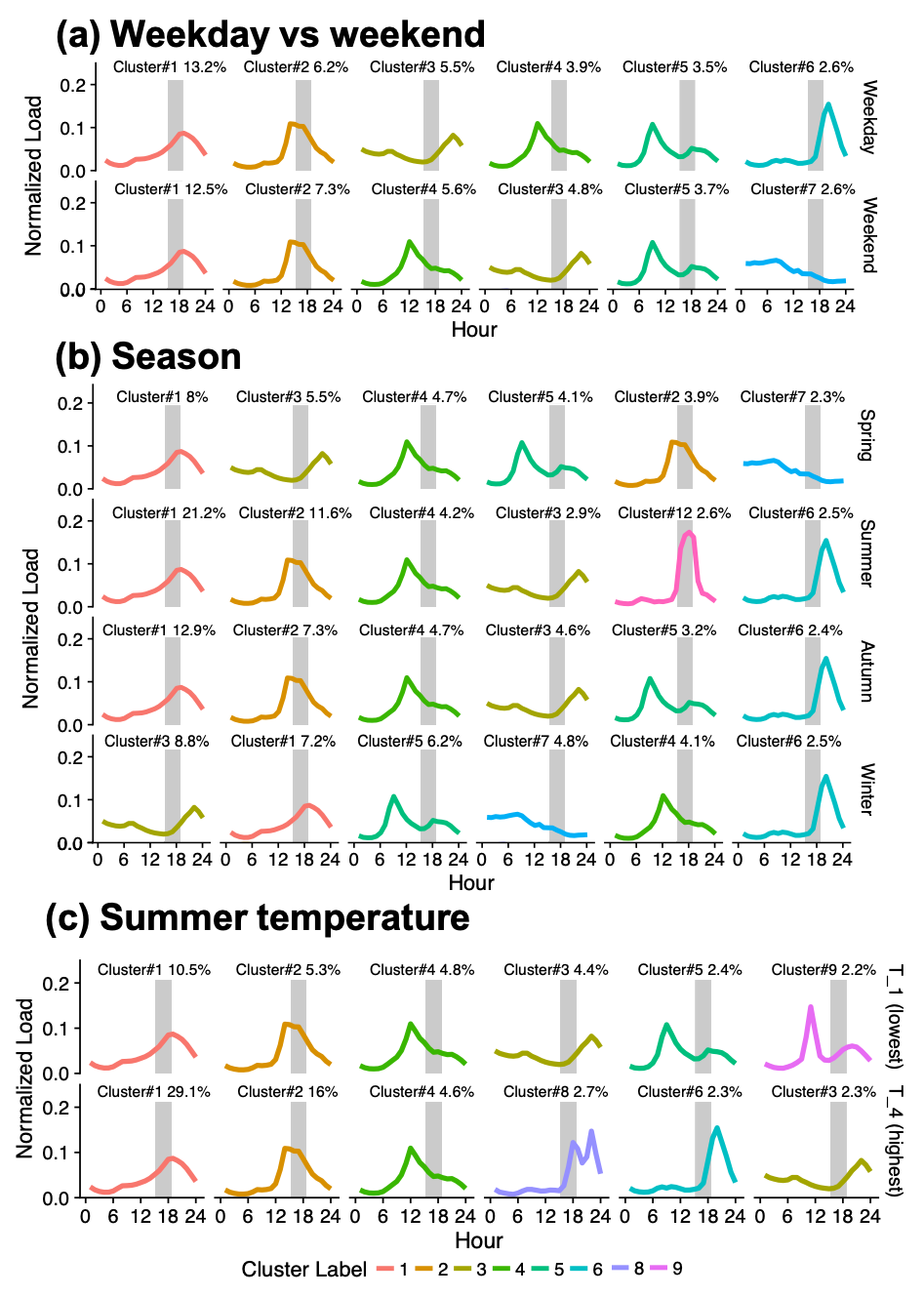}
  \caption{Top 6 dictionary load shapes and their occurrence frequency. (a) weekday vs weekend; (b) by season; and (c) by outdoor temperature in summer.}
  \label{fig:fig7}
  \end{center}
\end{figure}

\begin{figure}[ht]
\begin{center}
  \includegraphics[width=.88\columnwidth]{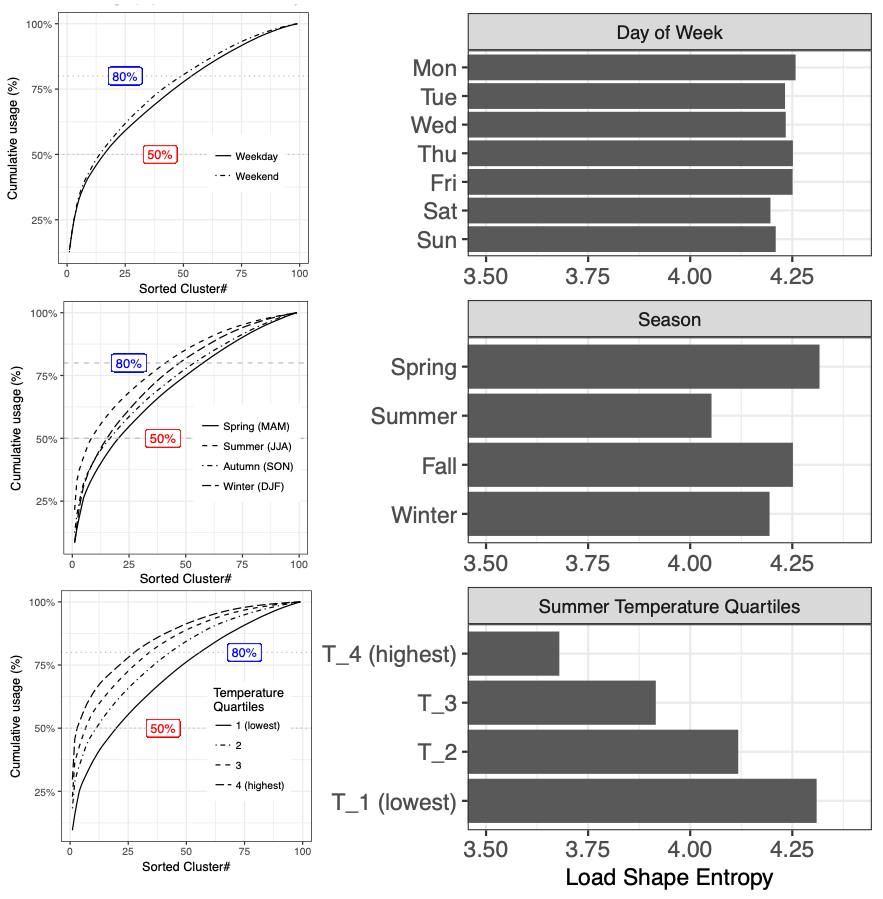}
  \caption{Cumulative electricity usage distribution of electricity usage across clusters (left column) and load shape entropy (right column) determined for temporal periods around external factors (day-of-week, season, and different levels of summer temperatures).}
  \label{fig:fig8}
  \end{center}
\end{figure}

This utility’s TOU rate is to be in effect for customers during the summer months only, and the higher peak price is only charged during the peak hours (4-7pm) on non-holiday weekdays. Because the season and days of the week are primary dimensions over which these rates are defined, we explore distributional differences in electricity usage patterns (Figure \ref{fig:fig7}) and their variability (Figure \ref{fig:fig8}) across these two timescales. In addition, we investigate the relationship of load shape patterns and their diversity with temperature in the summer season because increased consumption during peak hours on hot days represents the highest peak usage times for this utility. 

\subsubsection{Day-of-week variability}
most TOU rates, including the one offered by this utility, include peak period rates that are applicable only on weekdays. The representative usage patterns derived within and across households allow us to assess the degree to which weekdays differ from weekends with respect to the behavior patterns of these residential customers. According to Figure \ref{fig:fig7}a, the first observation of note is that five of these six dictionary load shapes are common between weekdays and weekends, while the 6th cluster differs. The 6th-ranked cluster on weekdays peaks in the evening, whereas the one on weekends peaks during the day, which indicates a slight increase of activities in the daytime at the residence on weekends.

In the left column of Figure \ref{fig:fig8} we compare the cumulative distributional differences between weekdays and weekends with respect to total electricity usage. From this figure we observe that weekend electricity consumption is slightly more concentrated in the top clusters (i.e., the weekend cumulative distribution is pulled slightly to the left of the weekday distribution), the entropy (right column of Figure \ref{fig:fig8}) suggests usage patterns on weekdays are relatively more diverse and variable than weekends. However, these differences are extremely small (difference in entropy is less than 0.07). These results suggest that at population level, the diversity of discretionary consumption schedules is not significantly informed by whether that usage is taking place on a weekend or a weekday. 

\subsubsection{Seasonal distributions}
according to the top 6 dictionary load shapes based on kWh usage coverage for each season (Figure \ref{fig:fig7}b), Cluster\#1, a shape with a single peak during the TOU period, is consistently one of the top two dictionary load shapes for all seasons and is the top dictionary load shape except for winter. In summer, cluster\#1 accounts for 21\% of summer electricity consumption of all households.  In winter, an evening peaking dictionary cluster (cluster\#2) covers ~9\% of the consumption while the kWh coverage of cluster\#1 drops to ~7\%. 

As can be seen in Figure \ref{fig:fig8}, electricity usage is most diverse (i.e. more uniformly distributed across the 99 dictionary load shapes) in the spring season and most concentrated and least variable in summer, with autumn and winter in between. Qualitatively, to see how significant this is, note that only 8 dictionary load shapes represent 50\% of the total summer electricity usage, while 20 dictionary load shapes are needed to cover the same proportion of total seasonal electricity consumption in winter. Cluster\#1 makes up almost a quarter of the summer electricity usage and is also the most populous cluster (15\% of total number of load shapes) in the summer. The difference in entropy between seasons (0.26) is much greater than day-of-week difference (less than 0.07).

 This suggests that factors necessary (be they socio-economic, preference-driven, or meteorological) to respond to summer conditions tend to concentrate and reduce the degree of variability in discretionary electricity usage patterns. It is a reasonable assumption, particularly given the high summer temperatures in this utility’s service territory, that this concentration in summer usage to a relatively small number of discretionary usage patterns is driven by the cooling needs and resulting air conditioning usage in the summer months\footnote{http://esnews.wapa.gov/wordpress/tag/sacramento-municipal-utility-district/}, which may be less variable across households than other types of usage. 

\subsubsection{Distribution by outdoor temperatures within the summer season} 
seasonal differences presented above indicate that this summer peaking utility has the least variable patterns during the peak demand season with significant cooling needs. Household energy consumption patterns in response to outdoor temperature itself is important to understand from the perspective of maintaining sufficient electricity generation capacity, grid reliability, and transmission and distribution infrastructure to meet demand. The most important factor is the demand on the small number of highest consumption days. Highest electricity consumption tends to occur on the hottest weekdays. 

We define daily average outdoor dry bulb temperature quartiles ( $<$68 F, 68-71F, 71-76F, and $>$76F) for the summer season. The top six dictionary load shapes (Figure \ref{fig:fig7}c) indicate that dictionary load shapes with peaks in the TOU period account for more electricity usage on the hottest days (''T\_4''), as compared to the cooler days of the summer. In particular, cluster\#1 peaks in the TOU period and represents 29\% of electricity consumed on the hottest days. By contrast, on the coolest summer days, the electricity consumed within this dominant dictionary cluster reduced by almost two-thirds (from 29.1\% to 10.5\% in Figure \ref{fig:fig7}c) . 

Usage patterns on the hottest days are more concentrated in the top shapes as indicated by the cumulative distribution and quantified by entropy (Figure \ref{fig:fig8}). There is a marked decreasing trend of entropy with temperature and the entropy difference between hottest and coolest periods is 0.63 (nearly three times seasonal difference in entropy, and ten times the day of week difference). The top 3 dictionary load shapes on the hottest days cover almost 50\% of the total electricity usage of the whole population during those days, while 18 dictionary load shapes are needed to cover the same proportion of usage during the coolest days. 

The key results from these analyses are: first, the degree to which behavioral patterns vary across seasons is much more significant than how they vary across weekends and weekdays. Second, seasonal variation in distributional differences is largely driven by temperature variation. Finally, given these results, we show that a relatively small number of usage patterns dominate the underlying drivers of energy consumption on the most important days when electricity demand is likely to put pressure on the grid (i.e., the hottest summer days). We show that three dictionary load shapes alone cover close to 50\% of the total energy consumption on the hottest summer days. This suggests that understanding the distribution of these three dictionary load shapes alone across households, and constructing outreach materials, messages, and programs around this information, could have a powerful impact on critical electricity load from the perspective of the grid. 

\subsection{Load shape variability within and cross customers}
\begin{figure}[ht]
\centering
  \includegraphics[width=.78\columnwidth]{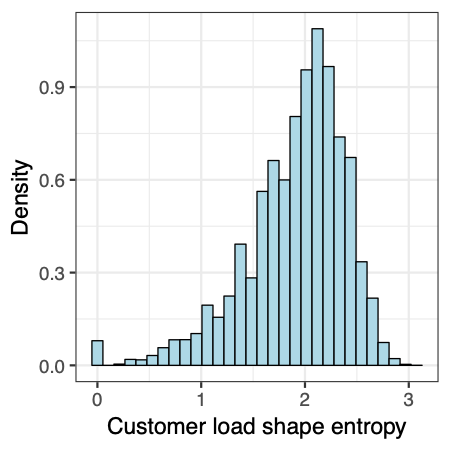}
  \caption{Distribution of summer load shape entropy of individual customers.}
  \label{fig:fig9}
\end{figure}

While overall consumption patterns are least diverse in the summer season, the variability in load shapes differ considerably among customers during this same period (Figure \ref{fig:fig9}). Entropy of individual customers ranges from 0 to 3, much greater than the overall difference driven by external factors such as day of week, season, or temperature, suggesting internal factors specific to individual households are more important to understand the behaviors that generate various degrees of load shape diversity.

\begin{figure}[ht]
\begin{center}
  \includegraphics[width=.98\columnwidth]{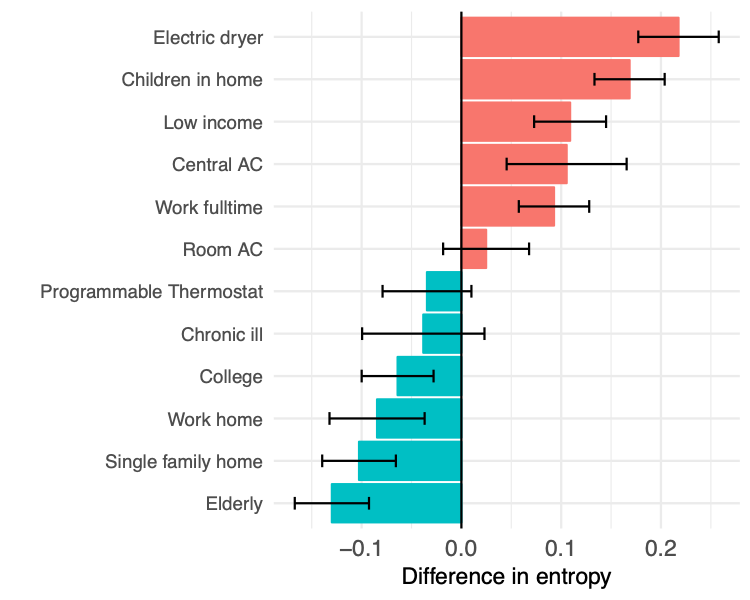}
  \caption{Difference in load shape entropy between households with and without the respective characteristic (error bars indicate 95\% confidence intervals).}
  \label{fig:fig12}
  \end{center}
\end{figure}

We assess the correlation between load shape entropy of individual households and household characteristics to understand the key predictors of variability in consumption schedules. Entropy differences between households with and without the selected household characteristics are shown in Figure \ref{fig:fig12}. We can see that having an electric dryer and children-in-home are the leading factors associated with more variable consumption schedules. This may be associated with variable day-to-day laundry time and more chaotic needs of children-in-home. Low-income households, those with central AC, and full-time workers also tend to vary their pattern of consumption from day to day. On the other end of the spectrum, elderly households tend to have the most stable day-to-day routines. Characteristics of single family homes, those working from home, and those with a college degree also tend to correlate with more stable consumption schedules relative to households without those characteristics. 

Depending on the household characteristics, we can see not all the resulting variability is likely to be readily translated to consumption flexibility. Usage of appliances such as electric dryers can be flexibly moved between times of day to accommodate a time-based DR program. However, the variability of load shapes due to having children-in-home reflects a lifestyle constraint that may be more difficult to change. 

\cite{economics2012demand} identified literature gaps with regard to consumption behavior and DR response potential of vulnerable and low-income customers. Our results suggest that in terms of variability in consumption schedules, households with chronic illness do not show significant differences relative to those without chronic illness. Low-income households on the other hand tend to be more variable while elderly households are the opposite. Further research is needed to test whether the variability shown in low-income households makes them desirable candidates for DR programs.


\begin{figure}[ht]
\begin{center}
  \includegraphics[width=.98\columnwidth]{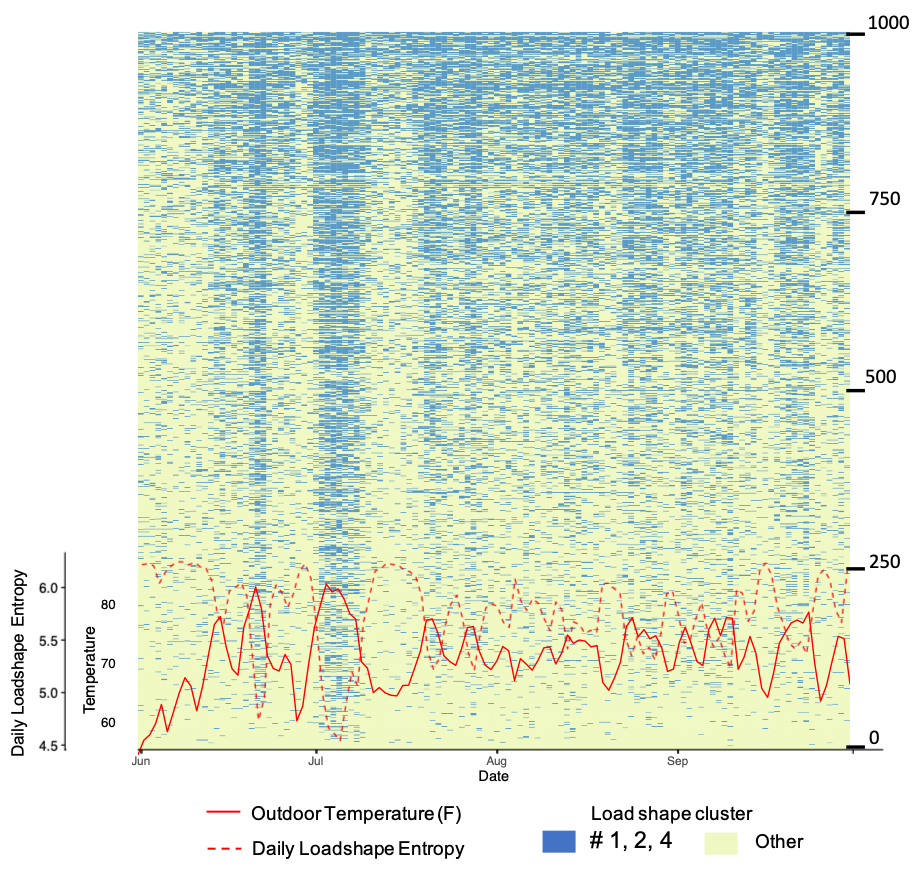}
  \caption{Occurrence of dictionary load shapes 1, 2 and 4 across households and relative to daily outdoor average temperature and load shape entropy.}
  \label{fig:fig11}
  \end{center}
\end{figure}

As discussed in the section IV B, there are three dictionary load shapes that cover 50\% of all the electricity consumption during the hottest summer days (clusters 1, 2 and 4 in Figure \ref{fig:fig7}c). To further understand the within and across customer variability we map the occurrence of these three dictionary load shapes of primary interest across the population of households in our data (a random sample of 1000 households are shown in Figure \ref{fig:fig11}). The households are sorted by the overall number of times one of their days was assigned to one of the three dictionary load shapes of interest. The households farthest to the bottom of the figure had the fewest days assigned to a cluster of interest, while those at the top had the most days assigned to a cluster of interest. The average daily outdoor temperature and load shape entropy computed for each day are overlaid onto the bottom of this figure for reference. 

In examining this figure, first the correlation of the clusters of interest with temperature is clear with counts of cluster-of-interest membership, i.e. column averages, increasing as temperatures spike. The diversity of usage patterns measured by load shape Entropy also shows sharp decreases when temperatures increase. Second, the variability in load shapes (whether in the top 3 dictionary patterns or not) can be explained by the responsiveness of the households to outside temperature, which is consistent with \cite{Fang2018-xc}. We see that the households in the top quarter of the figure are classified into one of the three target dictionary load shapes on a large percentage of their summer days, while the households in the bottom quarter of the figure only demonstrate these target usage patterns when temperatures spike. 

These two types of behavior, and the spectrum across households in between, represent very different underlying behavioral patterns. There is a need for further research to better understand these underlying patterns. Observable patterns such as these based on AMI data alone can be used to segment customers into groups that better capture heterogeneity in how they are likely to respond to DR or time-differentiated pricing programs. These segments could be used, for example, to target relevant programs based on the needs of the utility.

\section{Conclusion}
In this paper we employ an innovative clustering technique to categorize daily electricity consumption at hourly resolution across a large sample of residential customers over a full year. We focus clustering on the schedules and magnitudes of discretionary consumption with an innovative “de-minning” process. Our clustering procedure results in a dictionary of 99 distinctive usage patterns that can represent more than 30 million discretionary load shapes within a reasonable error threshold. 

With cluster assignment of daily discretionary load shapes over the whole year period, we are able to demonstrate how consumption patterns can be differentiated by external influencing factors such as time scales of interests (season and day-of-week) and meteorological conditions (outside temperature levels). Analysis of the temporal distribution of 99 dictionary load shapes reveals that high temperature is the single biggest external influence reducing discretionary load pattern diversity. In particular, approximately 50\% of the energy use during the hottest days are covered by only three dictionary load shapes. The coincident load resulting from increased concentration of usage patterns driven by high temperatures is problematic for the grid, causing high system peaks that are expensive and threaten service stability. Variation in the concentration of discretionary usage patterns exists across seasons as well, but this is largely being driven by temperature. There is much less variations in the distribution of electricity usage across dictionary load shapes between weekends and weekdays than across seasons. 

There is significant diversity of load shapes within households across days and such variability can be explained by household characteristics including socio-demographic and lifestyle information, dwelling information, and appliance ownership information. We find that having an electric dryer and children-in-home are the leading predictors of a more variable consumption schedule, while homes with elderly residents best predict stable day-to-day routines. Among the vulnerability characteristics considered here (chronic-illness, elderly, and low-income), we find low-income households tend to be more variable. This needs further research to confirm whether such variability can lead to greater DR potential.

We have demonstrated that there is significant heterogeneity across households regarding the diversity in usage patterns and such diversity markedly decreases on hotter days. The variability in summer load shapes across customers can be explained by the responsiveness of the households to outside temperature, which is consistent with current literature. Our results motivate future work in which identifying and mapping these particular patterns across the population can potentially be used in developing targeting techniques to improve the uptake and effectiveness of demand response programs.

While utilities and system operators typically focus on aggregate residential load shapes, our findings shed light on the considerable heterogeneity and the relative importance of influencing factors that underlying such variability across days and households. We argue that finding tractable ways to map out and understand this variability, as we have begun above, can be a powerful tool for subsequently segmenting customers for better program targeting and tailoring to meet the needs of the rapidly evolving electricity grid.


%



\section*{Acknowledgment}
The work described in this report was funded by Laboratory Directed Research and Development (LDRD) funds from the U.S. Department of Energy under Contract No. DE-AC02-05CH11231.

\ifCLASSOPTIONcaptionsoff
  \newpage
\fi



%



\bibliographystyle{IEEEtran}
\bibliography{jpapers}

\begin{thebibliography}{10}
\providecommand{\url}[1]{#1}
\csname url@samestyle\endcsname
\providecommand{\newblock}{\relax}
\providecommand{\bibinfo}[2]{#2}
\providecommand{\BIBentrySTDinterwordspacing}{\spaceskip=0pt\relax}
\providecommand{\BIBentryALTinterwordstretchfactor}{4}
\providecommand{\BIBentryALTinterwordspacing}{\spaceskip=\fontdimen2\font plus
\BIBentryALTinterwordstretchfactor\fontdimen3\font minus
  \fontdimen4\font\relax}
\providecommand{\BIBforeignlanguage}[2]{{%
\expandafter\ifx\csname l@#1\endcsname\relax
\typeout{** WARNING: IEEEtran.bst: No hyphenation pattern has been}%
\typeout{** loaded for the language `#1'. Using the pattern for}%
\typeout{** the default language instead.}%
\else
\language=\csname l@#1\endcsname
\fi
#2}}
\providecommand{\BIBdecl}{\relax}
\BIBdecl

\bibitem{Rasanen2010-xl}
T.~R{\"a}s{\"a}nen, D.~Voukantsis, H.~Niska, K.~Karatzas, and M.~Kolehmainen,
  ``Data-based method for creating electricity use load profiles using large
  amount of customer-specific hourly measured electricity use data,''
  \emph{Appl. Energy}, vol.~87, no.~11, pp. 3538--3545, Nov. 2010.

\bibitem{Flath2012-xl}
C.~Flath, D.~Nicolay, T.~Conte, C.~van Dinther, and L.~Filipova-Neumann,
  ``\BIBforeignlanguage{en}{Cluster analysis of smart metering data},''
  \emph{\BIBforeignlanguage{en}{Bus Inf Syst Eng}}, vol.~4, no.~1, pp. 31--39,
  Feb. 2012.

\bibitem{Cao2013-jz}
H.~Cao, C.~Beckel, and T.~Staake, ``Are domestic load profiles stable over
  time? an attempt to identify target households for demand side management
  campaigns,'' in \emph{{IECON} 2013 - 39th Annual Conference of the {IEEE}
  Industrial Electronics Society}.\hskip 1em plus 0.5em minus 0.4em\relax
  ieeexplore.ieee.org, Nov. 2013, pp. 4733--4738.

\bibitem{McLoughlin2015-eo}
F.~McLoughlin, A.~Duffy, and M.~Conlon, ``A clustering approach to domestic
  electricity load profile characterisation using smart metering data,''
  \emph{Appl. Energy}, vol. 141, pp. 190--199, Mar. 2015.

\bibitem{Khan2019-cz}
I.~Khan, M.~W. Jack, and J.~Stephenson, ``Identifying residential daily
  electricity-use profiles through time-segmented regression analysis,''
  \emph{Energy Build.}, vol. 194, pp. 232--246, Jul. 2019.

\bibitem{Kwac2014-rt}
J.~Kwac, J.~Flora, and R.~Rajagopal, ``Household energy consumption
  segmentation using hourly data,'' \emph{IEEE Trans. Smart Grid}, vol.~5,
  no.~1, pp. 420--430, Jan. 2014.

\bibitem{Haben2016-ma}
S.~Haben, C.~Singleton, and P.~Grindrod, ``Analysis and clustering of
  residential customers energy behavioral demand using smart meter data,''
  \emph{IEEE Trans. Smart Grid}, vol.~7, no.~1, pp. 136--144, Jan. 2016.

\bibitem{Yi_Wang2015-ff}
{Yi Wang}, {Qixin Chen}, {Chongqing Kang}, {Mingming Zhang}, {Ke Wang}, and
  {Yun Zhao}, ``Load profiling and its application to demand response: A
  review,'' \emph{Tsinghua Sci. Technol.}, vol.~20, no.~2, pp. 117--129, Apr.
  2015.

\bibitem{Wang2018-da}
Y.~Wang, Q.~Chen, T.~Hong, and C.~Kang, ``Review of smart meter data analytics:
  Applications, methodologies, and challenges,'' \emph{IEEE Transactions on
  Smart Grid}, Feb. 2018.

\bibitem{Chicco2012-ol}
G.~Chicco, ``Overview and performance assessment of the clustering methods for
  electrical load pattern grouping,'' \emph{Energy}, vol.~42, no.~1, pp.
  68--80, Jun. 2012.

\bibitem{Moslehi2010-bs}
K.~Moslehi and R.~Kumar, ``A reliability perspective of the smart grid,''
  \emph{IEEE Trans. Smart Grid}, 2010.

\bibitem{Farhangi2010-nx}
H.~Farhangi, ``The path of the smart grid,'' \emph{IEEE Power Energ. Mag.},
  vol.~8, no.~1, pp. 18--28, Jan. 2010.

\bibitem{Hong2011-au}
W.-C. Hong, ``Electric load forecasting by seasonal recurrent {SVR} (support
  vector regression) with chaotic artificial bee colony algorithm,''
  \emph{Energy}, vol.~36, no.~9, pp. 5568--5578, Sep. 2011.

\bibitem{Zhou2013-ya}
K.-L. Zhou, S.-L. Yang, and C.~Shen, ``A review of electric load classification
  in smart grid environment,'' \emph{Renewable Sustainable Energy Rev.},
  vol.~24, pp. 103--110, Aug. 2013.

\bibitem{Yilmaz2019-ra}
S.~Yilmaz, J.~Chambers, and M.~K. Patel, ``Comparison of clustering approaches
  for domestic electricity load profile characterisation - implications for
  demand side management,'' \emph{Energy}, vol. 180, pp. 665--677, Aug. 2019.

\bibitem{dyson2014using}
M.~E. Dyson, S.~D. Borgeson, M.~D. Tabone, and D.~S. Callaway, ``Using smart
  meter data to estimate demand response potential, with application to solar
  energy integration,'' \emph{Energy Policy}, vol.~73, pp. 607--619, 2014.

\bibitem{dent2014variability}
I.~Dent, T.~Craig, U.~Aickelin, and T.~Rodden, ``Variability of behaviour in
  electricity load profile clustering; who does things at the same time each
  day?'' in \emph{Industrial Conference on Data Mining}.\hskip 1em plus 0.5em
  minus 0.4em\relax Springer, 2014, pp. 70--84.

\bibitem{Public_Utilities_Commission2009-qu}
C.~Public Utilities~Commission, ``Order instituting rulemaking on the
  commission's own motion to consider alternative-fueled vehicle tariffs,
  infrastructure and policies to support …,'' \emph{California Public
  Utilities Commission}, 2009.

\bibitem{Jin2017-dv}
L.~Jin, D.~Lee, A.~Sim, S.~Borgeson, K.~Wu, C.~Anna~Spurlock, and A.~Todd,
  ``\BIBforeignlanguage{en}{Comparison of clustering techniques for residential
  energy behavior using smart meter data},'' in
  \emph{\BIBforeignlanguage{en}{Workshops at the {Thirty-First} {AAAI}
  Conference on Artificial Intelligence}}, Mar. 2017.

\bibitem{Dent2012-yu}
I.~Dent, T.~Craig, U.~Aickelin, and T.~Rodden, ``An approach for assessing
  clustering of households by electricity usage,'' Jan. 2012.

\bibitem{Milligan1988-ql}
G.~W. Milligan and M.~C. Cooper, ``A study of standardization of variables in
  cluster analysis,'' \emph{J. Classification}, vol.~5, no.~2, pp. 181--204,
  Sep. 1988.

\bibitem{Chicco2006-jz}
G.~Chicco, R.~Napoli, and F.~Piglione, ``Comparisons among clustering
  techniques for electricity customer classification,'' \emph{IEEE Trans. Power
  Syst.}, vol.~21, no.~2, pp. 933--940, May 2006.

\bibitem{Piao2014-gu}
M.~Piao, H.~S. Shon, J.~Y. Lee, and K.~H. Ryu, ``Subspace projection method
  based clustering analysis in load profiling,'' \emph{IEEE Trans. Power
  Syst.}, vol.~29, no.~6, pp. 2628--2635, Nov. 2014.

\bibitem{Han2011-ev}
J.~Han, J.~Pei, and M.~Kamber, \emph{\BIBforeignlanguage{en}{Data Mining:
  Concepts and Techniques}}.\hskip 1em plus 0.5em minus 0.4em\relax Elsevier,
  Jun. 2011.

\bibitem{torriti2015peak}
J.~Torriti, R.~Hanna, B.~Anderson, G.~Yeboah, and A.~Druckman, ``Peak
  residential electricity demand and social practices: Deriving flexibility and
  greenhouse gas intensities from time use and locational data,'' \emph{Indoor
  and Built Environment}, vol.~24, no.~7, pp. 891--912, 2015.

\bibitem{satre2019daily}
A.~Satre-Meloy, M.~Diakonova, and P.~Gr{\"u}newald, ``Daily life and demand: an
  analysis of intra-day variations in residential electricity consumption with
  time-use data,'' \emph{Energy Efficiency}, pp. 1--26, 2019.

\bibitem{Roberts2019-ck}
M.~B. Roberts, N.~Haghdadi, A.~Bruce, and I.~MacGill, ``Characterisation of
  australian apartment electricity demand and its implications for low-carbon
  cities,'' \emph{Energy}, May 2019.

\bibitem{Xu2017-lt}
S.~Xu, E.~Barbour, and M.~C. Gonz{\'a}lez, ``Household segmentation by load
  shape and daily consumption,'' in \emph{Proc. {ACM} {SigKDD} 2017 Conf.
  Halifax, Nov. Scotia, Canada, August 2017}.\hskip 1em plus 0.5em minus
  0.4em\relax urbcomp.ist.psu.edu, 2017.

\bibitem{Zhou2016-ec}
D.~Zhou, M.~Balandat, and C.~Tomlin, ``Residential demand response targeting
  using machine learning with observational data,'' Jul. 2016.

\bibitem{Fang2018-xc}
H.~Fang, Y.~Zhang, M.~Liu, and W.~Shen, ``Clustering and analysis of household
  power load based on {HMM} and multi-factors,'' in \emph{2018 {IEEE} 22nd
  International Conference on Computer Supported Cooperative Work in Design
  ({(CSCWD)})}.\hskip 1em plus 0.5em minus 0.4em\relax ieeexplore.ieee.org, May
  2018, pp. 491--495.

\bibitem{Chen2017-gy}
K.~Chen, J.~Hu, and Z.~He, ``Data-driven residential customer aggregation based
  on seasonal behavioral patterns,'' in \emph{2017 {IEEE} Power Energy Society
  General Meeting}.\hskip 1em plus 0.5em minus 0.4em\relax ieeexplore.ieee.org,
  Jul. 2017, pp. 1--5.

\bibitem{Melzi2017-re}
F.~N. Melzi, A.~Same, M.~H. Zayani, and L.~Oukhellou,
  ``\BIBforeignlanguage{en}{A dedicated mixture model for clustering smart
  meter data: Identification and analysis of electricity consumption
  behaviors},'' \emph{\BIBforeignlanguage{en}{Energies}}, vol.~10, no.~10, p.
  1446, Sep. 2017.

\bibitem{Rhodes2014-pu}
J.~D. Rhodes, W.~J. Cole, C.~R. Upshaw, T.~F. Edgar, and M.~E. Webber,
  ``Clustering analysis of residential electricity demand profiles,''
  \emph{Appl. Energy}, vol. 135, pp. 461--471, Dec. 2014.

\bibitem{economics2012demand}
F.~Economics and S.~First, ``Demand side response in the domestic sector-a
  literature review of major trials,'' \emph{Final Report, London, August},
  2012.

\bibitem{cappers2018vulnerable}
P.~Cappers, C.~A. Spurlock, A.~Todd, and L.~Jin, ``Are vulnerable customers any
  different than their peers when exposed to critical peak pricing: Evidence
  from the us,'' \emph{Energy policy}, vol. 123, pp. 421--432, 2018.

\bibitem{jin2016load}
L.~Jin, A.~Spurlock, S.~Borgeson, D.~Fredman, L.~Hans, S.~Patel, and A.~Todd,
  ``Load shape clustering using residential smart meter data: a technical
  memorandum,'' 2016.

\bibitem{Davies1979-wh}
D.~L. Davies and D.~W. Bouldin, ``\BIBforeignlanguage{en}{A cluster separation
  measure},'' \emph{\BIBforeignlanguage{en}{IEEE Trans. Pattern Anal. Mach.
  Intell.}}, vol.~1, no.~2, pp. 224--227, Feb. 1979.

\end{thebibliography}

%








\end{document}